\PassOptionsToPackage{unicode}{hyperref}
\PassOptionsToPackage{hyphens}{url}
\PassOptionsToPackage{dvipsnames,svgnames,x11names}{xcolor}
\documentclass[
]{article}
\usepackage{amsmath,amssymb}
\usepackage{lmodern}
\usepackage{iftex}
\ifPDFTeX
  \usepackage[T1]{fontenc}
  \usepackage[utf8]{inputenc}
  \usepackage{textcomp} 
\else 
  \usepackage{unicode-math}
  \defaultfontfeatures{Scale=MatchLowercase}
  \defaultfontfeatures[\rmfamily]{Ligatures=TeX,Scale=1}
\fi
\IfFileExists{upquote.sty}{\usepackage{upquote}}{}
\IfFileExists{microtype.sty}{
  \usepackage[]{microtype}
  \UseMicrotypeSet[protrusion]{basicmath} 
}{}
\makeatletter
\@ifundefined{KOMAClassName}{
  \IfFileExists{parskip.sty}{%
    \usepackage{parskip}
  }{
    \setlength{\parindent}{0pt}
    \setlength{\parskip}{6pt plus 2pt minus 1pt}}
}{
  \KOMAoptions{parskip=half}}
\makeatother
\usepackage{xcolor}
\newlength{\cslhangindent}
\newlength{\csllabelwidth}
\newlength{\cslentryspacingunit} 
\newenvironment{CSLReferences}[2] 
 {
  \setlength{\parindent}{0pt}
  \ifodd #1
  \let\oldpar\par
  \def\par{\hangindent=\cslhangindent\oldpar}
  \fi
  \setlength{\parskip}{#2\cslentryspacingunit}
 }%
 {}
\usepackage{calc}

\ifLuaTeX
\usepackage[bidi=basic]{babel}
\else
\usepackage[bidi=default]{babel}
\fi
\babelprovide[main,import]{american}

\def\languageshorthands#1{}
\ifLuaTeX
  \usepackage{selnolig}  
\fi
\IfFileExists{bookmark.sty}{\usepackage{bookmark}}{\usepackage{hyperref}}
\IfFileExists{xurl.sty}{\usepackage{xurl}}{} 
\hypersetup{
  pdftitle={binary\_c-python: A Python-based stellar population
synthesis tool and interface to binary\_c},
  pdfauthor={D. D. Hendriks, R. G. Izzard},
  pdflang={en-US},
  colorlinks=true,
  linkcolor={Maroon},
  filecolor={Maroon},
  citecolor={Blue},
  urlcolor={Blue},
  pdfcreator={LaTeX via pandoc}}

\title{\texttt{binary\_c-python}: A Python-based stellar population
synthesis tool and interface to \texttt{binary\_c}}

\usepackage[affil-it]{authblk}
\usepackage{orcidlink}
\author[1%
  ]{D. D. Hendriks%
    \,\orcidlink{0000-0002-8717-6046}\,%
      \thanks{E-mail:
\href{mailto:davidhendriks93@gmail.com}{\nolinkurl{davidhendriks93@gmail.com}}
(DDH)}%
  }
\author[1%
  ]{R. G. Izzard%
    \,\orcidlink{0000-0003-0378-4843}\,%
    }

\affil[1]{Department of Physics, University of Surrey, Guildford, GU2
7XH, Surrey, UK}
\date{11 April 2023}

\begin{document}
\maketitle

\hypertarget{summary}{%
\section{Summary}\label{summary}}

We present the software package
\href{https://binary_c.gitlab.io/binary_c-python/}{\texttt{binary\_c-python}}
which provides a convenient and easy-to-use interface to the
\href{https://binary_c.gitlab.io/binary_c}{\texttt{binary\_c}}
(\protect\hyperlink{ref-izzardNewSyntheticModel2004}{Izzard et al.,
2004},
\protect\hyperlink{ref-izzardPopulationNucleosynthesisSingle2006}{2006},
\protect\hyperlink{ref-izzardPopulationSynthesisBinary2009}{2009},
\protect\hyperlink{ref-izzardBinaryStarsGalactic2018}{2018};
\protect\hyperlink{ref-izzardCircumbinaryDiscsStellar2022}{Izzard \&
Jermyn, 2022}) framework, allowing the user to rapidly evolve individual
systems and populations of stars. \texttt{binary\_c-python} is available
on \href{https://pypi.org/project/binarycpython/}{\texttt{Pip}} and on
\href{https://binary_c.gitlab.io/binary_c-python/}{\texttt{GitLab}}.

\texttt{binary\_c-python} contains many useful features to control and
process the output of \texttt{binary\_c}, like by providing
\texttt{binary\_c-python} with logging statements that are dynamically
compiled and loaded into \texttt{binary\_c}. Moreover, we have recently
added standardised output of events like Roche-lobe overflow or double
compact-object formation to \texttt{binary\_c}, and automatic parsing
and managing of that output in \texttt{binary\_c-python}.
\texttt{binary\_c-python} uses multiprocessing to utilise all the cores
on a particular machine, and can run populations with HPC cluster
workload managers like \texttt{HTCondor} and \texttt{Slurm}, allowing
the user to run simulations on large computing clusters.

Recent developments in \texttt{binary\_c} include standardised output
datasets called \emph{ensembles}, which contain nested histograms of
binned data like supernovae rates or chemical yields.
\texttt{binary\_c-python} easily processes these datasets and provides a
suite of utility functions to handle them. Furthermore,
\texttt{binary\_c} now includes the \emph{ensemble-manager} class, which
makes use of the core functions and classes of \texttt{binary\_c-python}
to evolve a grid of stellar populations with varying input physics,
allowing for large, automated parameter studies through a single
interface.

\texttt{binary\_c-python} is easily integrated with other Python-based
codes and libraries, e.g.~sampling codes like \texttt{Emcee}
(\protect\hyperlink{ref-foreman-mackeyEmceeMCMCHammer2013}{Foreman-Mackey
et al., 2013}) or \texttt{Dynesty}
(\protect\hyperlink{ref-speagleDynestyDynamicNested2020}{Speagle,
2020}), or the astrophysics oriented package \texttt{Astropy}
(\protect\hyperlink{ref-astropycollaborationAstropyCommunityPython2013}{Astropy
Collaboration et al., 2013};
\protect\hyperlink{ref-astropycollaborationAstropyProjectBuilding2018}{Astropy
Collaboration et al., 2018}). Moreover, it is possible to provide custom
system-generating functions through function hooks, allowing third-party
packages to manage the properties of the stars in the populations and
evolve them through \texttt{binary\_c-python}.

We provide
\href{https://binary_c.gitlab.io/binary_c-python/readme_link.html}{documentation}
that is automatically generated based on \emph{docstrings} and a suite
of \texttt{Jupyter}
\href{https://binary_c.gitlab.io/binary_c-python/example_notebooks.html}{notebooks}.
These notebooks consist of technical tutorials on how to use
\texttt{binary\_c-python} and use-case scenarios aimed at doing science.
Much of \texttt{binary\_c-python} is covered by unit tests to ensure
reliability and correctness, and the test coverage is continually
increased as the package is improved.

\hypertarget{statement-of-need}{%
\section{Statement of need}\label{statement-of-need}}

In the current scientific climate \texttt{Python} is ubiquitous. While
lower level code written in, e.g., \texttt{Fortran} or \texttt{C} is
still widely used, much new software is written in \texttt{Python},
either entirely or as a wrapper around other code and libraries.
Education in programming also often includes \texttt{Python} courses
because of its ease of use and its flexibility. Moreover,
\texttt{Python} has a large community with many resources and tutorials.
We have created \texttt{binary\_c-python} to allow students and
scientists alike to explore current scientific issues while enjoying the
familiar syntax, and at the same time make use of the plentiful
scientific and astrophysical packages like \texttt{Numpy},
\texttt{Scipy}, \texttt{Pandas}, \texttt{Astropy} and platforms like
\texttt{Jupyter}.

Over time, many (binary) stellar population synthesis codes have been
created. Each one usually has a slightly different focus, like
gravitational waves or spectral synthesis. Recent ones are
\texttt{BPASS/HOKI} (\protect\hyperlink{ref-hoki}{Stevance et al.,
2020}), \texttt{COMPAS} (\protect\hyperlink{ref-COMPAS}{Compas et al.,
2022}), \texttt{COSMIC} (\protect\hyperlink{ref-COSMIC}{Breivik et al.,
2019}), \texttt{MOBSE} (\protect\hyperlink{ref-MOBSE}{Giacobbo \&
Mapelli, 2019}) and \texttt{SeBa} (\protect\hyperlink{ref-seba}{Toonen
\& Nelemans, 2013}). Most of these have \texttt{Python} interfaces,
which further highlights the need to develop and release a modern
\texttt{Python} interface to \texttt{binary\_c},
i.e.~\texttt{binary\_c-python}.

The previous interface to \texttt{binary\_c}, \texttt{binary\_grid} was
written in \texttt{Perl}, where much of the logic and structure was
developed and debugged. While much of the code base of
\texttt{binary\_c-python} has changed significantly from its
\texttt{Perl} predecessor, the initial porting to \texttt{Python} and
the development of \texttt{binary\_c-python} greatly benefitted from the
existence of this earlier interface.

\hypertarget{projects-that-use-binary_c-python}{%
\section{\texorpdfstring{Projects that use
\texttt{binary\_c-python}}{Projects that use binary\_c-python}}\label{projects-that-use-binary_c-python}}

\texttt{binary\_c-python} has already been used in a variety of
situations, ranging from pure research to educational purposes, as well
as in outreach events. In the summer of 2021 we used
\texttt{binary\_c-python} as the basis for the interactive classes on
stellar ecosystems during the
\href{https://www2.mpia-hd.mpg.de/imprs-hd/SummerSchools/2021/}{International
Max-Planck Research School summer school 2021 in Heidelberg}. Students
were introduced to the topic of population synthesis and were able to
use our notebooks to perform their own calculations.
\texttt{binary\_c-python} has been used in Mirouh et al.
(\protect\hyperlink{ref-mirouh_etal22}{submitted}), implementing
improvements to tidal interactions between stars and varying initial
birth parameter distributions to match to observed binary systems in
star clusters. The \texttt{binary\_c-python} and \texttt{Emcee} packages
were used in a Master's thesis project to find the birth system
parameters of the V106 stellar system, compare observations to results
of \texttt{binary\_c}, and perform uncertainty with Bayesian uncertainty
inference through Markov chain Monte Carlo sampling.

Currently, \texttt{binary\_c-python} is used in several ongoing
projects. These include a study on the effect of birth distributions on
the occurrence of carbon-enhanced metal-poor (CEMP) stars, the occurence
and properties of accretion disks in main-sequence stars
(\protect\hyperlink{ref-hendriks_etal2023_disks}{Hendriks \& Izzard, in
prep}), and the predicted observable black hole distribution by
combining star formation and metallicity distributions with the output
of \texttt{binary\_c}
(\protect\hyperlink{ref-hendriks_etal2023_gw}{Hendriks et al., in
prep}). We also use the \emph{ensemble} output structure to generate
datasets for galactochemical evolution over cosmological timescales,
where we rely heavily on the utilities of \texttt{binary\_c-python}
(\protect\hyperlink{ref-yates_etal2023}{Yates et al., in prep}).

\hypertarget{acknowledgements}{%
\section{Acknowledgements}\label{acknowledgements}}

We are grateful for the helpful discussions and testing efforts from A.
Aryeaipour, M. Delorme, S. Dykes, G. Mirouh, M. Matteuzzi, N. Rees, M.
Renzo, L. van Son, K. Temmink, D. Tracey and R. Yates, and the early
work of J. Andrews which inspired our Python-C interface code. DDH
thanks the UKRI/UoS for the funding grant H120341A. RGI thanks STFC for
funding grants
\href{https://gtr.ukri.org/projects?ref=ST\%2FR000603\%2F1}{ST/R000603/1}
and \href{https://gtr.ukri.org/projects?ref=ST/L003910/2}{ST/L003910/2}.

\hypertarget{references}{%
\section*{References}\label{references}}
\addcontentsline{toc}{section}{References}

\hypertarget{refs}{}
\begin{CSLReferences}{1}{0}
\leavevmode\vadjust pre{\hypertarget{ref-astropycollaborationAstropyProjectBuilding2018}{}}%
Astropy Collaboration, Price-Whelan, A. M., Sipőcz, B. M., Günther, H.
M., Lim, P. L., Crawford, S. M., Conseil, S., Shupe, D. L., Craig, M.
W., Dencheva, N., Ginsburg, A., VanderPlas, J. T., Bradley, L. D.,
Pérez-Suárez, D., de Val-Borro, M., Aldcroft, T. L., Cruz, K. L.,
Robitaille, T. P., Tollerud, E. J., \ldots{} Astropy Contributors.
(2018). The {Astropy Project}: {Building} an {Open-science Project} and
{Status} of the v2.0 {Core Package}. \emph{The Astronomical Journal},
\emph{156}, 123. \url{https://doi.org/10.3847/1538-3881/aabc4f}

\leavevmode\vadjust pre{\hypertarget{ref-astropycollaborationAstropyCommunityPython2013}{}}%
Astropy Collaboration, Robitaille, T. P., Tollerud, E. J., Greenfield,
P., Droettboom, M., Bray, E., Aldcroft, T., Davis, M., Ginsburg, A.,
Price-Whelan, A. M., Kerzendorf, W. E., Conley, A., Crighton, N.,
Barbary, K., Muna, D., Ferguson, H., Grollier, F., Parikh, M. M., Nair,
P. H., \ldots{} Streicher, O. (2013). Astropy: {A} community {Python}
package for astronomy. \emph{Astronomy and Astrophysics}, \emph{558},
A33. \url{https://doi.org/10.1051/0004-6361/201322068}

\leavevmode\vadjust pre{\hypertarget{ref-COSMIC}{}}%
Breivik, K., Coughlin, S., Zevin, M., Rodriguez, C., Andrews, J.,
Kimball, C., mcdigman, \& 1nhtran. (2019).
\emph{{COSMIC-PopSynth/COSMIC: COSMIC now integrated with Globular
Cluster Simulation Software ClusterMonteCarlo (CMC)}} (Version v3.1.0).
Zenodo. \url{https://doi.org/10.5281/zenodo.3482915}

\leavevmode\vadjust pre{\hypertarget{ref-COMPAS}{}}%
Compas, T., Riley, J., Agrawal, P., Barrett, J. W., Boyett, K. N. k.,
Broekgaarden, F. S., Chattopadhyay, D., Gaebel, S. M., Gittins, F.,
Hirai, R., Howitt, G., Justham, S., Khandelwal, L., Kummer, F., Lau, M.
Y. m., Mandel, I., Mink, S. E. de, Neijssel, C., Riley, T., \ldots{}
Willcox, R. (2022). COMPAS: A rapid binary population synthesis suite.
\emph{Journal of Open Source Software}, \emph{7}(69), 3838.
\url{https://doi.org/10.21105/joss.03838}

\leavevmode\vadjust pre{\hypertarget{ref-foreman-mackeyEmceeMCMCHammer2013}{}}%
Foreman-Mackey, D., Hogg, D. W., Lang, D., \& Goodman, J. (2013). Emcee:
{The MCMC Hammer}. \emph{Publications of the Astronomical Society of the
Pacific}, \emph{125}(925), 306--312.
\url{https://doi.org/10.1086/670067}

\leavevmode\vadjust pre{\hypertarget{ref-MOBSE}{}}%
Giacobbo, N., \& Mapelli, M. (2019). {The impact of electron-capture
supernovae on merging double neutron stars}. \emph{482}(2), 2234--2243.
\url{https://doi.org/10.1093/mnras/sty2848}

\leavevmode\vadjust pre{\hypertarget{ref-hendriks_etal2023_disks}{}}%
Hendriks, D. D., \& Izzard, R. G. (in prep). {Disk mass loss during mass
transfer onto main sequence accretors}.

\leavevmode\vadjust pre{\hypertarget{ref-hendriks_etal2023_gw}{}}%
Hendriks, D. D., van Son, L. A. C., Renzo, M., \& Izzard, R. G. (in
prep). {Pulsational pair-instability supernovae in gravitational-wave
and electromagnetic transients}.

\leavevmode\vadjust pre{\hypertarget{ref-izzardPopulationNucleosynthesisSingle2006}{}}%
Izzard, R. G., Dray, L. M., Karakas, A. I., Lugaro, M., \& Tout, C. A.
(2006). {Population nucleosynthesis in single and binary stars. I.
Model}. \emph{460}(2), 565--572.
\url{https://doi.org/10.1051/0004-6361:20066129}

\leavevmode\vadjust pre{\hypertarget{ref-izzardPopulationSynthesisBinary2009}{}}%
Izzard, R. G., Glebbeek, E., Stancliffe, R. J., \& Pols, O. R. (2009).
{Population synthesis of binary carbon-enhanced metal-poor stars}.
\emph{508}(3), 1359--1374.
\url{https://doi.org/10.1051/0004-6361/200912827}

\leavevmode\vadjust pre{\hypertarget{ref-izzardCircumbinaryDiscsStellar2022}{}}%
Izzard, R. G., \& Jermyn, A. S. (2022). Circumbinary discs for stellar
population models. \emph{Monthly Notices of the Royal Astronomical
Society}. \url{https://doi.org/10.1093/mnras/stac2899}

\leavevmode\vadjust pre{\hypertarget{ref-izzardBinaryStarsGalactic2018}{}}%
Izzard, R. G., Preece, H., Jofre, P., Halabi, G. M., Masseron, T., \&
Tout, C. A. (2018). {Binary stars in the Galactic thick disc}.
\emph{473}(3), 2984--2999. \url{https://doi.org/10.1093/mnras/stx2355}

\leavevmode\vadjust pre{\hypertarget{ref-izzardNewSyntheticModel2004}{}}%
Izzard, R. G., Tout, C. A., Karakas, A. I., \& Pols, O. R. (2004). {A
new synthetic model for asymptotic giant branch stars}. \emph{350}(2),
407--426. \url{https://doi.org/10.1111/j.1365-2966.2004.07446.x}

\leavevmode\vadjust pre{\hypertarget{ref-mirouh_etal22}{}}%
Mirouh, G. M., Hendriks, D. D., Dykes, S., Moe, M., \& Izzard, R. G.
(submitted). {Detailed equilibrium and dynamical tides: impact on
circularization and synchronization in open clusters}.

\leavevmode\vadjust pre{\hypertarget{ref-speagleDynestyDynamicNested2020}{}}%
Speagle, J. S. (2020). Dynesty: A dynamic nested sampling package for
estimating {Bayesian} posteriors and evidences. \emph{Monthly Notices of
the Royal Astronomical Society}, \emph{493}(3), 3132--3158.
\url{https://doi.org/10.1093/mnras/staa278}

\leavevmode\vadjust pre{\hypertarget{ref-hoki}{}}%
Stevance, H., Eldridge, J., \& Stanway, E. (2020). {Hoki: Making BPASS
accessible through Python}. \emph{The Journal of Open Source Software},
\emph{5}(45), 1987. \url{https://doi.org/10.21105/joss.01987}

\leavevmode\vadjust pre{\hypertarget{ref-seba}{}}%
Toonen, S., \& Nelemans, G. (2013). {The effect of common-envelope
evolution on the visible population of post-common-envelope binaries}.
\emph{557}, A87. \url{https://doi.org/10.1051/0004-6361/201321753}

\leavevmode\vadjust pre{\hypertarget{ref-yates_etal2023}{}}%
Yates, R. M., Hendriks, D. D., Vijayan, A. P., Izzard, R. G., Thomas, P.
A., \& Das, P. (in prep). {Modelling binary stellar populations \& dust
in cosmological-scale galaxy evolution simulations}.

\end{CSLReferences}

\end{document}